\input harvmac
\sequentialequations
\overfullrule=0pt

\def\dmunphys{d\mu^{(n)}_{\rm phys}}
\def\dmuncl{d\mu^{(n)}_{\rm cl}}
\def\dmutwophys{d\mu^{(2)}_{\rm phys}}
\def\barvhiggsa{\bar{\cal A}_{\sst00}}

\def\VolOn{{\rm Vol}\big(O(n)\big)}

\def\Pinfty{{\cal P}_{\sst\infty}}
\def\Atot{{\A_{\rm tot}}}

\def\uA{\,\lower 1.2ex\hbox{$\sim$}\mkern-13.5mu A}
\def\bigL{{\bf L}}

\def\zero{{\scriptscriptstyle(0)}}
\font\authorfont=cmcsc10 \ifx\answ\bigans\else scaled\magstep1\fi
\divide\baselineskip by 10
\multiply\baselineskip by 9
\def\prenomat{\hbox{hep\hbox{-}th/9709072}}
\Title{$\prenomat$}{\vbox{\centerline{Supersymmetry and the
Multi-Instanton Measure}\centerline{}
\centerline{II. From $N=4$ to $N=0$ }}
}
\centerline{\authorfont Nicholas Dorey \rm and \authorfont Timothy J. 
Hollowood}
\bigskip
\centerline{\sl Physics Department, University of Wales Swansea,
 Swansea SA2$\,$8PP UK}
\centerline{\tt n.dorey@swansea.ac.uk$\,,$ t.hollowood@swansea.ac.uk}
\bigskip
\centerline{\authorfont Valentin V. Khoze}
\bigskip
\centerline{\sl Department of Physics, Centre for Particle Theory, 
University of Durham}
\centerline{\sl Durham DH1$\,$3LE UK $\quad$ \tt valya.khoze@durham.ac.uk}
\bigskip
\centerline{and}
\bigskip
\centerline{\authorfont Michael P. Mattis}
\bigskip
\centerline{\sl Theoretical Division T-8, Los Alamos National Laboratory}
\centerline{\sl Los Alamos, NM 87545 USA$\quad$ \tt mattis@pion.lanl.gov}
\vskip .3in
\def\hf{{\textstyle{1\over2}}}

\def\quarter{{\textstyle{1\over4}}}
\noindent
Extending recent $N=1$ and $N=2$ results, we propose an
explicit formula for the integration measure on the moduli
space of \hbox{charge-$n$} ADHM multi-instantons in $N=4$
supersymmetric $SU(2)$ gauge theory. As a consistency check,
we derive a renormalization group
relation between the $N=4,$ $N=2$, and $N=1$ measures. We then use this
relation
to construct the purely bosonic (``$N=0$'') measure as well, in the classical
approximation in which the one-loop
small-fluctuations determinants is not included.
\vskip .1in
\Date{\bf September 1997 } 
\vfil\break

\lref\ADS{I. Affleck, Nucl. Phys. B191 (1981) 429;
 I. Affleck, M. Dine and N. Seiberg, Nucl. Phys. B241
(1984) 493; Nucl. Phys. B256 (1985) 557.  }
\lref\RGorig{S. Weinberg, Phys. Lett. 91B (1980) 51;
L. Hall, Nucl. Phys. B178 (1981) 75.}
\lref\RGinst{D. Finnell and P. Pouliot,
{\it Instanton calculations versus exact results in 4 dimensional 
SUSY gauge theories},
Nucl. Phys. B453 (95) 225, hep-th/9503115;
N. Dorey, V.V. Khoze and M.P. Mattis, \it On $N=2$
supersymmetric QCD with $4$ flavors\rm,  hep-th/9611016, 
Nucl.~Phys.~\rm B492 \rm (1997) 607, Sec.~2.}
\lref\detrefsO{E. Corrigan, P. Goddard, H. Osborn and S. Templeton,
Nucl.~Phys.~B159 (1979) 469; H. Osborn, Nucl.~Phys.~B159 (1979) 497;
H. Osborn and G. P. Moody, Nucl.~Phys.~B173 (1980) 422;
I. Jack, Nucl.~Phys.~B174 (1980) 526;
I. Jack and H. Osborn, Nucl.~Phys.~B207 (1982) 474.}
\lref\detrefsL{H. Berg and M. Luscher, Nucl.~Phys.~B160 (1979) 281;
H. Berg and J. Stehr, Nucl.~Phys.~B173 (1980) 422 and B175 (1980) 293.}
\lref\AH{M. Atiyah and N. Hitchin, {\it ``The Geometry and Dynamics
of Magnetic Monopoles''}, Princeton University Press (1988).}
\lref\gm{G. Gibbons and N. Manton, {\rm Nucl. Phys.} {\rm B274} (1986) 183.}
\lref\dkmten{N. Dorey, V.V. Khoze and M.P. Mattis, \it
Supersymmetry and the multi-instanton measure\rm, hep-th/9708036.}
\lref\dadda{A. D'Adda and P. Di Vecchia, 
{\rm Phys. Lett.} {\rm 73B} (1978) 162.}
\lref\GMO{P. Goddard, P. Mansfield, and H. Osborn,  Phys.~Lett.~\rm98B
\rm (1981) 59.}
\lref\NSVZ{ V. A. Novikov, M. A. Shifman, A. I. Vainshtein and
V. I. Zakharov, Nucl Phys. B229 (1983) 394; Nucl. Phys. B229 (1983)
407; Nucl. Phys. B260 (1985) 157. }
\lref\ADHM{M.  Atiyah, V.  Drinfeld, N.  Hitchin and
Yu.~Manin, Phys. Lett. A65 (1978) 185. }
\lref\Osborn{H. Osborn, Ann. Phys. 135 (1981) 373. }
\lref\CGTone{ E. Corrigan, P. Goddard and S. Templeton,
Nucl. Phys. B151 (1979) 93; 
   E. Corrigan, D. Fairlie, P. Goddard and S. Templeton,
    Nucl. Phys. B140 (1978) 31.}
\lref\dkmone{N. Dorey, V.V. Khoze and M.P. Mattis, \it Multi-instanton
calculus in $N=2$ supersymmetric gauge theory\rm, hep-th/9603136,
Phys.~Rev.~D54 (1996) 2921.}
\lref\dkmfive{N. Dorey, V.V. Khoze and M.P. Mattis, \it On $N=2$
supersymmetric QCD with $4$ flavors\rm,  hep-th/9611016, 
Nucl.~Phys.~\bf B492 \rm (1997) 607.}
\lref\dkmsix{N. Dorey, V.V. Khoze and M.P. Mattis, \it On mass-deformed $N=4$
supersymmetric Yang-Mills theory\rm,  hep-th/9612231, 
Phys.~Lett.~B396 (1997) 141.}
\lref\dkmfour{N. Dorey, V.V. Khoze and M.P. Mattis, \it Multi-instanton
calculus in $N=2$ supersymmetric gauge theory.
II. Coupling to matter\rm, hep-th/9607202, Phys.~Rev.~D54 (1996) 7832.}
\lref\tHooft{G. 't Hooft, Phys. Rev. D14 (1976) 3432; ibid.
D18 (1978) 2199.}
\def\frac#1#2{{ {#1}\over{#2}}}

\def\trtwo{\tr^{}_2\,}

\def\abar{\bar a}

\def\dalpha{{\dot\alpha}}

\def\sst{\scriptscriptstyle}

\def\A{{\cal A}}
\def\susy{supersymmetry}

\def\cl{{\,\rm cl}}
\def\lambdabar{\bar\lambda}

\def\psibar{\bar\psi}
\def\sqrtwo{\sqrt{2}\,}

\def\Qbar{\bar Q}
\def\susic{supersymmetric}

\def\vhiggsa{{{\cal A}_{\sst00}}}

\def\C{{\cal C}}

\def\new{{\scriptscriptstyle\rm new}}

\def\uX{\,\lower 1.2ex\hbox{$\sim$}\mkern-13.5mu X}
\def\uQ{\,\lower 1.2ex\hbox{$\sim$}\mkern-13.5mu Q}
\def\uQtilde{\,\lower 1.2ex\hbox{$\sim$}\mkern-13.5mu \tilde Q}
\def\uD{\,\lower 1.2ex\hbox{$\sim$}\mkern-13.5mu {\rm D}}

\def\uF{\,\lower 1.2ex\hbox{$\sim$}\mkern-13.5mu F}
\def\uW{\,\lower 1.2ex\hbox{$\sim$}\mkern-13.5mu W}
\def\uWbar{\,\lower 1.2ex\hbox{$\sim$}\mkern-13.5mu {\overline W}}
\def\uPhibar{\,\lower 1.2ex\hbox{$\sim$}\mkern-13.5mu {\overline \Phi}}

\def\uV{\,\lower 1.2ex\hbox{$\sim$}\mkern-13.5mu V}
\def\uv{\,\lower 1.0ex\hbox{$\scriptstyle\sim$}\mkern-11.0mu v}
\def\uPsi{\,\lower 1.2ex\hbox{$\sim$}\mkern-13.5mu \Psi}
\def\uPhi{\,\lower 1.2ex\hbox{$\sim$}\mkern-13.5mu \Phi}
\def\uchi{\,\lower 1.5ex\hbox{$\sim$}\mkern-13.5mu \chi}
\def\uchitilde{\,\lower 1.5ex\hbox{$\sim$}\mkern-13.5mu \tilde\chi}
\def\Psibar{\bar\Psi}
\def\uPsibar{\,\lower 1.2ex\hbox{$\sim$}\mkern-13.5mu \Psibar}
\def\upsi{\,\lower 1.5ex\hbox{$\sim$}\mkern-13.5mu \psi}
\def\uq{\,\lower 1.5ex\hbox{$\sim$}\mkern-13.5mu q}
\def\uqtilde{\,\lower 1.5ex\hbox{$\sim$}\mkern-13.5mu \tilde q}
\def\psibar{\bar\psi}
\def\upsibar{\,\lower 1.5ex\hbox{$\sim$}\mkern-13.5mu \psibar}
\def\upsibarzero{\,\lower 1.5ex\hbox{$\sim$}\mkern-13.5mu \psibar^\zero}
\def\ulambda{\,\lower 1.2ex\hbox{$\sim$}\mkern-13.5mu \lambda}
\def\ulambdabar{\,\lower 1.2ex\hbox{$\sim$}\mkern-13.5mu \lambdabar}
\def\ulambdabarzero{\,\lower 1.2ex\hbox{$\sim$}\mkern-13.5mu \lambdabar^\zero}
\def\ulambdabarnew{\,\lower 1.2ex\hbox{$\sim$}\mkern-13.5mu \lambdabar^\new}
\def\D{{\cal D}}
\def\M{{\cal M}}
\def\N{{\cal N}}
\def\Dslash{\,\,{\raise.15ex\hbox{/}\mkern-12mu \D}}
\def\Dbarslash{\,\,{\raise.15ex\hbox{/}\mkern-12mu {\bar\D}}}
\def\delslash{\,\,{\raise.15ex\hbox{/}\mkern-9mu \partial}}
\def\delbarslash{\,\,{\raise.15ex\hbox{/}\mkern-9mu {\bar\partial}}}

\def\hf{{\textstyle{1\over2}}}
\def\quarter{{\textstyle{1\over4}}}
\def\eighth{{\textstyle{1\over8}}}

\def\xibar{\bar\xi}

\def\uAcl{\,\lower 1.2ex\hbox{$\sim$}\mkern-13.5mu A^{}_{\cl}}
\def\uAbarcl{\,\lower 1.2ex\hbox{$\sim$}\mkern-13.5mu A_{\cl}^\dagger}

\def\deltafcn{\hbox{$\delta$-function}}
\def\deltafcns{\hbox{$\delta$-functions}}
\newsec{Introduction}
In a recent paper \dkmten\ (henceforth (I)$\,$) we constructed
the collective coordinate integration measure for charge-$n$ ADHM
multi-instantons \ADHM\ in both $N=1$ and $N=2$ \susic\ $SU(2)$ gauge
theories. Here we present the analogous formula for the $N=4$ 
theory. As a nontrivial consistency check, 
we  derive a renormalization group (RG) relation between
the $N=4$ and $N=2$ measures, and between the $N=2$ and $N=1$ measures,
that emerges when the appropriate components of the supermultiplets are
given a mass which is taken to infinity. 
In turn, this RG relation also yields an interesting formula
for the purely bosonic (``$N=0$'') ADHM measure. However,
unlike the \susic\ cases, this $N=0$ measure is valid at the classical
level only, i.e., excluding the one-loop 
small-fluctuations \hbox{'t Hooft} determinants
over positive-frequency gauge and ghost modes in the self-dual 
background \tHooft. 
(It is not necessary to invoke
this ``classical approximation'' in the $N=1,2,4$ cases,
as the 't Hooft determinants  cancel  between bosonic and fermionic
excitations so long as there is at least one supersymmetry \dadda.)
Since  substantial progress has been made towards the
calculation of these determinants in the ADHM background
\refs{\detrefsO,\detrefsL}, there is reason
for optimism that our field-theoretic understanding of the
multi-instanton sector
in the $N=0$ model
will come to match our current understanding of the single-instanton sector.

In what follows we will focus on pure $N=0,1,2$ or 4 \susic\
 gauge theories; the incorporation of additional
matter in the fundamental representation of the gauge
group is straightforward (see Sec.~4 of (I)$\,$). 
For general topological number $n$, the $N=1$ collective coordinate
integration measure $\dmunphys$ is given in 
Eqs.~(2.23) and (2.54) of (I):\foot{The ADHM and SUSY notation and conventions
are as in (I). In particular $(ij)_n$ and $\langle ij\rangle_n$ stand
for the ordered pairs $(i,j)$ subject to $1\le i\le j\le n$ and
$1\le i<j\le n,$ respectively. Also see (I) for references to the
earlier literature.}
\def\ijn{{(ij)^{}_n}}

\def\ijna{{\langle ij\rangle^{}_n}}

\eqn\dmudef{\eqalign{\int\dmunphys\ &=\ 
{(C_1)^n\over\VolOn}\int\prod_{i=1}^nd^4w_id^2\mu_i
\prod_{\ijn}d^4a'_{ij}d^2\M'_{ij}
\cr&\times\ \prod_{\ijna}\prod_{c=1,2,3}
\delta\big(\quarter\trtwo\tau^c[(\abar a)_{i,j}-
(\abar a)_{j,i}]\big)\,\delta^2\big((\abar \M)_{i,j}-
(\abar \M)_{j,i}\big)\ .}}
Here
\eqn\bcanonical{a_{\alpha\dalpha}\ =\  
\pmatrix{w_{1\alpha\dalpha}&\cdots&w_{n\alpha\dalpha}
\cr{}&{}&{}\cr
{}&a'_{\alpha\dalpha}&{}\cr{}&{}&{}}\quad,\qquad
\M^\gamma\ =\ \pmatrix{\mu_1^\gamma&\cdots&\mu_n^\gamma
\cr{}&{}&{}\cr
{}&\M^{\prime\gamma}&{}\cr{}&{}&{}}}
are, respectively, $(n+1)\times n$  quaternion-valued and
Weyl-spinor-valued collective coordinate matrices describing $8n$ 
independent
bosonic (gauge field) and $4n$ independent 
fermionic (gaugino) degrees of freedom
of the super-multi-instanton. These matrices
are subject to the symmetry conditions
$a'=a^{\prime T}$ and $\M'=\M^{\prime T}$ as well as to the supersymmetrized
ADHM constraints \refs{\ADHM,\CGTone}
implemented by the \deltafcns\ in \dmudef\ (which are
absent in the single-instanton sector, $n=1$). Also $C_1$ is 't Hooft's
 1-instanton factor \tHooft
\eqn\Coneequals{C_1\ =\ 2^9\,
\Lambda_{\sst N=1}^6\ \propto\ \exp(-8\pi^2/g^2_{\sst N=1})}
where $\Lambda_{\sst N=1}$ is the dynamically generated scale in the
Pauli-Villars (PV) scheme, which is the natural scheme for instanton
calculations \tHooft.

Since the \deltafcns\ in Eq.~\dmudef\ are dictated by the ADHM formalism,
and since, as shown in (I), the resulting measure turns out to be
 a \susy\ invariant and also has the correct transformation property under
the anomalous $U(1)_{R}$ symmetry,
we made the stronger claim in (I) that this Ansatz is in fact unique.
To see why, let us consider including an additional
function of the collective coordinates, $f(a,\M),$ in the integrand
of Eq.~\dmudef. To preserve \susy, we can require that $f$
be a \susy\ invariant.  It is a fact that 
any non-constant function that is a \susy\ invariant
must contain fermion bilinear pieces (and possibly higher powers of
fermions as well). By the rules of Grassmann integration, such bilinears
would necessarily
lift some of the adjoint fermion zero modes contained in $\M.$ But since
Eq.~\dmudef\ contains precisely the right number
of unlifted fermion zero modes dictated by the $U(1)_{R}$ anomaly, 
namely $4n$, this argument rules
out the existence of a non-constant function $f$. Moreover, any
constant $f$ would be absorbed into the overall
multiplicative factor, which is fixed by cluster decomposition
as detailed in (I). In fact, a similar uniqueness argument applies
to our proposed
ADHM measure for $N=2$ theories. Unfortunately, as we
discuss below, the
above argument cannot be applied directly to the $N=4$ model where
the anomaly vanishes. 

In the $N=2$ model the story is slightly more
complicated due to the presence of an adjoint scalar field. However, in the
absence of a VEV for this field, the anomaly dictates a total of $8n$
unlifted modes. The appropriate measure can then be determined by the
same considerations as  for the $N=1$ case, including the above
uniqueness argument. The $N=2$
formula is given in Eqs.~(3.19) and (3.27) of (I):
\eqn\dmudeftwo{\eqalign{\int\dmunphys\ 
&=\ {(C'_1)^n\over\VolOn}\int\prod_{i=1}^nd^4w_id^2\mu_i d^2\nu_i
\prod_{\ijn}d^4a'_{ij}d^2\M'_{ij}d^2\N'_{ij}
\cr&\times\ \prod_{\ijna}\prod_{c=1,2,3}
\delta\big(\quarter\trtwo\tau^c[(\abar a)_{i,j}-
(\abar a)_{j,i}]\big)\,\delta^2\big((\abar \M)_{i,j}-(\abar \M)_{j,i}\big)
\cr&\qquad\qquad\qquad
\times\ \delta^2\big((\abar \N)_{i,j}-(\abar \N)_{j,i}\big)\,
(\det\bigL)^{-1}
 .}}
Here $\N,$ like $\M$, is a Weyl-spinor-valued matrix containing $4n$
independent adjoint Higgsino degrees of freedom; $\bigL$ is a certain
$\hf n(n-1)\times\hf n(n-1)$ linear operator on the space of $n\times n$
antisymmetric matrices (see Eq.~(3.10) of (I)$\,$); and the 1-instanton
factor is \tHooft
\eqn\Conepequals{C'_1\ =\ 2^8\,\pi^{-4}\,
\Lambda_{\sst N=2}^4\ \propto\ \exp(-8\pi^2/g^2_{\sst N=2})\ }
with $\Lambda_{\sst N=2}$ in the PV scheme as before. 

The main application for the above measure is in the calculation of
instanton corrections to the Coulomb branch of the $N=2$ theory, where
the VEV of the adjoint scalar spontaneously breaks the gauge
group down to an abelian subgroup.  In the presence of a VEV, the 
$U(1)_{R}$ symmetry is also spontaneously broken which spoils the
naive fermion zero mode counting implied by the anomaly. In addition, the
instanton is no longer an exact solution of the equations of motion.  
These two features lead to an instanton action which depends explicitly 
on bosonic and fermionic collective coordinates, the latter dependence
lifting all but four of the $8n$ fermion zero modes.  
This case is analyzed in detail in \dkmone\ using the constrained
instanton of Affleck, Dine and Seiberg \ADS. An important feature
of the constrained instanton approach is that, at leading semiclassical order,
the effects of a non-zero VEV enter {\it only} through the instanton 
action and hence there are no additional modifications to the measure (4).    
The relevant multi-instanton
actions for a variety of $N=2$ \susic\ models are assembled in Sec.~5
of (I). 

\newsec{Ansatz for the $N=4$ measure} Next we turn to the
$N=4$ case, which is reviewed in Ref.~\dkmsix. In studying this model
 it is convenient to relabel
$\M\rightarrow\M^1$ and $\N\rightarrow\M^2$ as in \dkmsix. The $N=4$ model
requires two additional adjoint fermion multiplets (adjoint Higgsinos), 
parametrized by collective coordinate matrices
$\M^3$ and $\M^4.$ An $SU(4)_R$ symmetry acts on these superscripts.
The multi-instanton action for $N=4$ \susic\ $SU(2)$ gauge theory then
reads \dkmsix:
\def\Lambdabar{\bar\Lambda}

\eqn\fouraction{\eqalign{S_{\rm inst}^{\sst N=4}\ &=\ 
 16\pi^2|\vhiggsa|^2\sum_{k=1}^n|w_k|^2
\ -\ 8\pi^2\,\Tr_n\big(\A'\cdot\Lambdabar+\A'_{f}(\M^1,\M^2)\cdot\Lambdabar
-\A'_{f}(\M^3,\M^4)\cdot\Lambda\big)
\cr&\ \ +\   
4\sqrtwo\pi^2\,\sum_{k=1}^n\,
\big(\mu_k^{1\alpha}\,\barvhiggsa{}_\alpha{}^\beta\,
\mu^2_{k\beta}\ -\ \mu_k^{3\alpha}\,\vhiggsa{}_\alpha{}^\beta\,\mu^4_{k\beta}
\big)\cr&\ \ 
+\ \pi^2\sum_{A,B,C,D=1}^4\ \epsilon_{ABCD}\Tr_n\,\A_f'(\M^A,\M^B)\cdot
\Lambda_f(\M^C,\M^D)\  .}}
Here $\vhiggsa$ is the $SU(2)$-valued VEV, which we have chosen to live
in the Higgs which is the $N=1$ superpartner of the $\M^1$ Higgsino;
 the collective coordinates
 $w_k$ and $\mu_k^A$ are as in Eq.~\bcanonical;
and $\Lambda$ and $\Lambda_f(\M^A,\M^B)$ are as in (I). Also
$\A'$ and $\A'_f(\M^A,\M^B)$ are defined as the solutions to
$\bigL\cdot\A'=\Lambda$ and $\bigL\cdot\A_f'(\M^A,\M^B)=\Lambda_f(\M^A,
\M^B)$. 

We now discuss the $N=4$ collective coordinate integration measure. As in the
$N=1$ and $N=2$ cases we seek an Ansatz with the following 
 four properties:\hfil\break\indent
(i) Invariance under $N=4$ \susy; \hfil\break\indent
(ii) Invariance under the internal $O(n)$ transformations which are redundant
degrees of freedom endemic to the ADHM construction; \hfil\break\indent
(iii) cluster decomposition in the dilute-gas limit  of large
space-time separation between instantons; \hfil\break\indent   
(iv) agreement with known formulae in the 1-instanton sector.
\hfil\break\noindent
We will show that the following expression embodies these properties:
\eqn\dmudefour{\eqalign{\int\dmunphys\ 
&=\ {(C''_1)^n\over\VolOn}\int\prod_{A=1,2,3,4}\,
\prod_{i=1}^nd^4w_id^2\mu^A_i 
\prod_{\ijn}d^4a'_{ij}d^2\M^{A\prime}_{ij}
\cr&\times\ \prod_{\ijna}\prod_{c=1,2,3}
\delta\big(\quarter\trtwo\tau^c[(\abar a)_{i,j}-
(\abar a)_{j,i}]\big)\,\prod_{A=1,2,3,4}\,
\delta^2\big((\abar \M^A)_{i,j}-(\abar \M^A)_{j,i}\big)
\cr&\qquad\qquad\qquad
\times\ (\det\bigL)^{-3}\ 
 ,}}
where $C_1''$ is the 1-instanton factor, again in the PV scheme \tHooft:
\eqn\Coneppequals{C''_1\ =\ 2^6\,\pi^{-12}\,
 \exp(-8\pi^2/g^2_{\sst N=4})\ .}
Note that there is no dynamically generated scale in the $N=4$ model as it
is conformally invariant, and finite (albeit scheme dependent due to
cancellations between individually divergent diagrams).
In the 1-instanton sector the second and third lines of Eq.~\dmudefour\
are omitted, as in the $N=1$ and $N=2$ cases.

Before verifying properties (i)-(iv), we remind the reader that the
$N=1$ and $N=2$ measures also needed to satisfy a fifth defining property
discussed in (I):
the number of adjoint fermion zero modes left unsaturated
by the measure had to equal $4n$ and $8n$, respectively. This fermionic
mode counting is dictated by the anomaly, and is at the heart of the
uniqueness argument given above. But in the $N=4$ model the anomaly
vanishes, and the issue of fermionic mode counting is less definite.
To see this ambiguity, note that
 the Ansatz \dmudefour\ leaves $16n$ Grassmann modes
unlifted, i.e., twice the $N=2$ counting as one might expect. On the
other hand it appears to us to be
 purely a matter of convention whether or not the
exponentiated action \fouraction\ should be considered part of the measure,
especially in the limit of vanishing VEV.
And here the $N=4$ action differs in a significant way
from its $N=1$ and $N=2$ counterparts: in this limit
the action remains nontrivial, specifically the last term in Eq.~\fouraction\
survives. If one then chooses to include this fermion quadrilinear term
\eqn\ifinclude{
\exp\big(- \pi^2\sum_{A,B,C,D=1}^4\ \epsilon_{ABCD}\Tr_n\,\A_f'(\M^A,\M^B)\cdot
\Lambda_f(\M^C,\M^D)\,\big) }
in the definition of the $N=4$ measure, then the number of unlifted
Grassmann modes falls from $16n$ to 16. As discussed in \dkmsix,
these 16 are precisely the modes generated by the 
the eight supersymmetric and 
eight superconformal symmetries of the Lagrangian which are broken
by the instanton.

Despite the disappearance of the anomaly in the $N=4$ case, the
remaining properties
(i)-(iv) are highly restrictive, and we believe the Ansatz
\dmudefour\ to be unique.
As  further nontrivial checks, we will also compare  this proposed measure
against the known first-principles expression 
in the 2-instanton sector \refs{\Osborn,\dkmone}.
Moreover we will derive, and verify, a stringent RG relation between the
measures \dmudef, \dmudeftwo\ and \dmudefour. This will also serve to relate
the 1-instanton factors $C^{}_1,$ $C_1'$ and $C_1''.$

\newsec{Supersymmetric invariance of the $N=4$ measure}
The proof of properties (ii), (iii), and (iv) proceeds just as for the
$N=1$ and $N=2$ cases in (I), and need not be repeated here. In particular,
the cluster condition (iii) fixes the overall $n$-instanton
constant in \dmudefour\ in
terms of the 1-instanton factor $C''_1,$  as before. Here
we need only
focus on property (i), invariance under $N=4$ \susy. Let us recall how
this was established in (I) in the $N=1$ and $N=2$ cases. Under an
infinitesimal $N=1$ \susy\ transformation $\xibar\Qbar+\xi Q,$ one has
\refs{\dkmfour,\NSVZ}
$\delta a_{\alpha\dalpha}=\xibar_\dalpha\M_\alpha$ and
 $\delta\M_\alpha=-4ib\xi_\alpha,$ with $b$ as in (I). 
Thus the argument of the second \deltafcn\ in \dmudef\ is invariant, while
the argument of the first \deltafcn\ picks up an admixture of the second.
It follows that the product of the \deltafcns\ is an $N=1$ invariant.

Next we turn to the $N=2$ measure \dmudeftwo. The trick here is to represent
$(\det\bigL)^{-1}$ as an integral:\foot{Here, and in the $N=2$ and $N=4$
\susy\ algebras to follow, we set the adjoint Higgs VEV to zero for
simplicity. This
suffices for the collective coordinate integration measure, which is
necessarily the same with or without a VEV as noted above. 
When the VEV vanishes,
$\Atot(\M,\N)\equiv\A'+\A'_f(\M,\N)$ collapses to the fermion bilinear
$\A'_f(\M,\N).$}
\eqn\detLrep{(\det\bigL)^{-1}\ =\ 
\int\prod_{\ijna}d\Atot(\M,\N)_{i,j}\ 
\delta\big(\,(\bigL\cdot\Atot(\M,\N)
-\Lambda_f(\M,\N))_{i,j}\,\big)\ 
 ,}
with $\Atot$ and $\Lambda_f$ as in (I).
Now consider the behavior of the arguments of the four \deltafcns,
respectively the three in Eq.~\dmudeftwo\ and the one in Eq.~\detLrep,
under an infinitesimal $N=2$ transformation 
$\xibar_1\Qbar_1+\xibar_2\Qbar_2+\xi_1Q_1+\xi_2Q_2.$
Recall the action of the $N=2$ \susy\ on the collective coordinates \dkmfour:
\eqna\susyalgebratwo
$$\eqalignno{
&\delta a_{\alpha\dalpha}\ =\ \xibar_{1\dalpha}\M_\alpha
+\xibar_{2\dalpha}\N_\alpha \qquad&\susyalgebratwo a \cr
&\delta\M_\gamma\ =\ -4ib\xi_{1\gamma}
-2\sqrtwo\C_{\gamma\dalpha}(\M,\N)\,
\xibar_2^\dalpha  \qquad&\susyalgebratwo b \cr
&\delta\N_\gamma\ =\ -4ib\xi_{2\gamma}+2\sqrtwo \C_{\gamma\dalpha}(\M,\N)\,
\xibar_1^\dalpha \qquad&\susyalgebratwo c \cr
&\delta\Atot(\M,\N)\ =\ 0\ \qquad&\susyalgebratwo d \cr
}$$
Here $\C_{\gamma\dalpha}(\M,\N)$ 
is the $(n+1)\times n$ quaternion-valued matrix
\eqn\Cdef{\C(\M,\N)\ =\
\pmatrix{-w_k\Atot(\M,\N)_{k1}&\cdots&
-w_k\Atot(\M,\N)_{kn}
\cr
{}&{}&{}\cr
{}&\big[\,\Atot(\M,\N)\,,\,a'\,]&{}\cr
{}&{}&{}  }\ .}
Thus $\Atot(\M,\N),$ in addition to being the dummy of integration in
Eq.~\detLrep, also completes the $N=2$ algebra; see (I) for a review of its
relation to the adjoint Higgs. Using Eq.~\susyalgebratwo{},
 it is straightforward to show that,
 under $\xibar_1\Qbar_1,$ the arguments of the four \deltafcns\ transform
into linear combinations of one another, as follows. The argument of the
second 
\deltafcn\ is invariant; the arguments of the first and fourth
pick up an admixture
of the second; and the argument of the third picks up an admixture of the
fourth. (Under $\xibar_2\Qbar_2,$ exchange the roles of the second and
third \deltafcns; also $\xi_1Q_1$ and $\xi_2Q_2$ act trivially.) 
Clearly the superdeterminant of such an
``upper triangular'' linear
transformation is unity; it follows that the product of the \deltafcns\ is
indeed an $N=2$ invariant.

Lastly we turn to the $N=4$ measure, Eq.~\dmudefour. We wish to demonstrate
that it is invariant under an $N=4$ \susy\ transformation
$\sum_{A=1}^4\xibar_A\Qbar_A+\xi_AQ_A.$ 
The obvious extension of Eq.~\susyalgebratwo{}
to the $N=4$ case, consistent with $SU(4)_R$ symmetry, reads:
\eqna\susyalgebrafour
$$\eqalignno{
&\delta a_{\alpha\dalpha}\ =\ \sum_{A=1,2,3,4}\,
\xibar_{A\dalpha}\M^A_\alpha
&\susyalgebrafour a \cr
&\delta\M^A_\gamma\ =\ -4ib\xi_{A\gamma}-2\sqrtwo\sum_{B=1,2,3,4}\,
\C_{\gamma\dalpha}(\M^A,\M^B)\,
\xibar_B^\dalpha  &\susyalgebrafour b \cr
&\delta_A\Atot(\M^A,\M^B)\ =\ \delta_B\Atot(\M^A,\M^B)\ =\ 
0\ &\susyalgebrafour c \cr
}$$
where $\delta_A$ denotes a variation under $\xibar_A\Qbar_A+\xi_AQ_A$
 only (no sum on $A$). 
Here we are using the fact that in the absence of a VEV both $\Atot$ and
$\C$ are antisymmetric in their fermionic arguments: $\C(\M^A,\M^B)=
-\C(\M^B,\M^A)$ and $\C(\M^A,\M^A)=0.$ Note that Eq.~\susyalgebrafour b
encompasses not only Eqs.~\susyalgebratwo{b,c} given above, but
also the $N=2$ transformation law for the adjoint Higgsinos $\M^3$ and
$\M^4$; see Eq.~(12) of Ref.~\dkmsix. The $N=4$ algebra \susyalgebrafour{}
is completed by giving the (nonvanishing) transformation law for
 $\delta_C\Atot(\M^A,\M^B)$ where $A,$ $B$ and $C$ are distinct. This
cumbersome expression is straightforward to derive from the defining
equation for $\Atot$ \dkmone,
but is not actually needed below. Finally we note that
this $N=4$ algebra, like the $N=2$ algebra \susyalgebratwo{}, 
is correct
only equivariantly, up to transformations by the internal $O(n)$ 
group \dkmfour.
Consequently it should only be applied to $O(n)$ singlets, which suffices
for present purposes (see property (ii) above). 

Using Eq.~\susyalgebrafour{}, 
we can now check that the proposed measure \dmudefour\ is indeed an $N=4$
invariant. For concreteness let us focus (say) on the fourth supersymmetry,
$\xibar_4\Qbar_4$. As in the $N=2$ case, we introduce an integral
representation for $(\det\bigL)^{-3}$:
\eqn\detLrepa{(\det\bigL)^{-3}\ =\ 
\int\prod_{A\neq 4}\prod_{\ijna}d\Atot(\M^A,\M^4)_{i,j}\ 
\delta\big(\,(\bigL\cdot\Atot(\M^A,\M^4)
-\Lambda_f(\M^A,\M^4))_{i,j}\,\big)\ 
 .}
The index $A$ ranges over the three supersymmetries orthogonal to the
one under examination, in this case $A=1,2,3.$ Under
$\xibar_4\Qbar_4$, the arguments of the first \deltafcn\ in Eq.~\dmudefour\
and of the three \deltafcns\ in Eq.~\detLrepa\
gain an admixture of the fermionic constraint
$(\abar \M^4)_{i,j}-(\abar \M^4)_{j,i}\,$, which is itself invariant as per
Eq.~\susyalgebrafour b. Likewise the arguments of the remaining three
fermionic \deltafcns, namely
$(\abar \M^A)_{i,j}-(\abar \M^A)_{j,i}$ with $A=1,2,3,$ gain admixtures
of the arguments of the three corresponding \deltafcns\ in
\detLrepa.
So, once again, the  linear transformation has an upper-triangular
structure with
superdeterminant
unity, implying that the
product of all the \deltafcns\ in Eqs.~\dmudefour\ and
\detLrepa\ is invariant under
$\xibar_4\Qbar_4$. Invariance of the measure \dmudefour\
under the other three $\xibar_A\Qbar_A$
follows by permuting the indices in the above discussion, whereas the
$\xi_AQ_A$ act trivially as before.

\def\Jfermi{J_{\rm fermi}}
\newsec{Two-instanton check of the $N=4$ measure}
As a first nontrivial consistency check of our proposed $N=4$ measure
\dmudefour, let us show that it agrees with the known first-principles
measure in the 2-instanton sector \refs{\Osborn,\dkmone}. The discussion
exactly parallels that of Sec.~3.5 of (I) for the $N=2$ case, except
that the fermionic zero-mode Jacobian $\Jfermi$ should be squared,
there being twice as many adjoint fermion zero modes in the $N=4$ model.
Consequently one has \refs{\Osborn,\dkmone}:
\def\Stwo{{\cal S}_2}

\eqn\dmuOsb{\eqalign{\int\dmutwophys\ =\ {(C_1'')^2
\over\Stwo}\int &d^4w_1d^4w_2
d^4a'_{11}d^4a'_{22}\prod_{A=1,2,3,4}\,d^2\mu^A_1d^2\mu^A_2
d^2\M^{A\prime}_{11}d^2\M^{A\prime}_{22}
\cr&\times\ {64|a_3|^4\over H^3}\
\Big|\,|a_3|^2
\,-\,|a'_{12}|^2\,-\,\eighth{d\Sigma^\phi
\over d\phi}\big|_{\phi=0}\,\Big|\ ,}}
in the notation of (I).
Like Eqs.~(2.55) and (3.28) in (I), this is a ``gauge fixed'' measure, the
form of which explicitly breaks both $O(2)$ and \susy\ invariance. Also
as in (I),
the overall factor in Eq.~\dmuOsb\ is tied to 't Hooft's 1-instanton factor
$C_1''$ by cluster decomposition.\foot{In the clustering limit taken
in (I), $H\rightarrow 4|a_3|^2\rightarrow\infty$,
 which accounts for the factor of 64 in
Eq.~\dmuOsb. Similarly, the right-hand side of Eq.~(3.29) of (I) should
contain a factor of 4.}
Inserting the factors of unity
\eqn\unityb{1\, =\, 16|a_3|^4\int d^4a'_{12}\,
\prod_{c=1,2,3}
\delta\big(\quarter\trtwo\tau^c[(\abar a)_{1,2}-(\abar a)_{2,1}]\big)
\delta\big(\abar_3^{}a'_{12}+\bar a'_{12}a^{}_3-\hf
\Sigma(a_3,a_0,w_1,w_2)\big)}
and
\eqn\unityf{1\ =\ \prod_{A=1,2,3,4}\ {1\over4|a_3|^2}\int d^2
\M^{A\prime}_{12}\,
\delta^2\big((\abar\M^A)_{1,2}-(\abar\M^A)_{2,1}\big)}
into Eq.~\dmuOsb,
performing the change of dummy integration variables $a\rightarrow
a^\phi$ and $\M^A\rightarrow \M^{A\phi}$ described
 in (I),  inserting a final factor of unity
$1 = (2\pi)^{-1}\int_0^{2\pi}d\phi$, 
and recalling that in the 2-instanton sector $\det\bigL=H,$
one readily recovers the $O(2)$- and $N=4$ invariant form for the measure,
Eq.~\dmudefour. See Secs.~2.5 and 3.5 of (I) for calculational details.

\newsec{RG relation between the $N=1$, $N=2$, and $N=4$ measures}
A distinct consistency check is to invoke the physical requirement of 
RG decoupling to relate the $N=1,$ $N=2$ and $N=4$ measures. Let us
focus first on the $N=1$ and $N=2$ measures, Eqs.~\dmudef\ and \dmudeftwo.
In the language of $N=1$ superfields, the particle content of $N=2$
\susic\ Yang-Mills theory consists of a gauge superfield $V=
(v_m,\lambda_\alpha)$ and an adjoint chiral superfield $\Phi=(A,\psi_\alpha).$
Let us add a mass term $m\,\trtwo\Phi^2\big|_{\theta^2}+\hbox{H.c.}$ for the
matter superfield, breaking the $N=2$ \susy\ down to $N=1.$ To leading
semiclassical approximation, this is equivalent to inserting a Higgsino
mass factor \dkmsix
\eqn\Higgsmass{\exp\big(-m\pi^2\,\Tr_n\,\N^{\gamma T}(\Pinfty+1)\N_\gamma\big)}
into the integrand of Eq.~\dmudeftwo. Here $\Pinfty$ is the
$(n+1)\times(n+1)$ matrix diag$(1,0,\cdots,0)$ in the conventions of 
\Osborn.  (There are bosonic mass terms too
but their effect on the semiclassical physics is
 down by one factor of the coupling as they require the
elimination of an auxiliary $F$ field.) RG decoupling means that in the double
scaling limit for the mass and coupling constant, 
defined by $m\rightarrow\infty$
and $g\rightarrow0$ with a certain combination held fixed (namely, the
left-hand side of Eq.~(23b) below),
 the $\Phi$ multiplet decouples from the physics. Concomitantly,
the $N=2$ measure should collapse to the $N=1$ measure. Comparing
Eqs.~\dmudef\ and \dmudeftwo\ then leads to the RG consistency requirement
in this limit:
\eqn\RGreq{\eqalign{&(C'_1)^n\int\prod_{i=1}^n d^2\nu_i
\prod_{\ijn}d^2\N'_{ij}
\prod_{\ijna}\,
\delta^2\big((\abar \N)_{i,j}-(\abar \N)_{j,i}\big)\,
\exp\big(-m\pi^2\,\Tr_n\,\N^{\gamma T}(\Pinfty+1)\N_\gamma\big)
\cr&\qquad\qquad\longrightarrow\ 
(C_1)^n\cdot\det\bigL\ .}}

Note that each side of Eq.~\RGreq\ is a complicated function of the bosonic
moduli $a$, and it is far from obvious that the two sides are proportional
to one another. We therefore regard this relation as a stringent
consistency check on our proposed integration measures.
To establish this proportionality, we examine the integral
\def\I{{\cal I}}
\eqn\Indef{\eqalign{\I_n(a;m)\ &=\ 
\int\prod_{i=1}^n d^2\nu_i
\prod_{\ijn}d^2\N'_{ij}
\prod_{\ijna}d\Atot(\M,\N)_{i,j}
\cr&\times\ 
\prod_{\ijna}\,
\delta^2\big((\abar \N)_{i,j}-(\abar \N)_{j,i}\big)\,
\delta\big(\,(\bigL\cdot\Atot(\M,\N)
-\Lambda_f(\M,\N))_{i,j}\,\big)\cr&\times\ 
\exp\big(-m\pi^2\,\Tr_n\,\N^{\gamma T}(\Pinfty+1)\N_\gamma\big)\ .}}
On the one hand, by construction,
$\I_n$ is a function of the bosonic collective coordinates
$a$ only. On the other hand, from Eq.~\susyalgebratwo{}, one confirms
that $\I_n$ is an $N=1$  invariant under $\xibar_1\Qbar_1+\xi_1Q_1.$\foot{This
claim is only true if $\M$ satisfies the super-ADHM relation $\abar\M=
(\abar\M)^T$ \CGTone, but this suffices for our needs, since this
relation is enforced by \deltafcns\ in both Eqs.~\dmudef\ and \dmudeftwo.}
 But, as mentioned earlier,
 the only purely bosonic \susy\ invariants are constants! Thus $\I_n(a;m)
\equiv\I_n(m),$ independent of the matrix $a$. Performing
the $\Atot$ integration in Eq.~\Indef\ using Eq.~\detLrep, and
comparing to Eq.~\RGreq, establishes the claimed proportionality, and
leads to the condition
\eqn\Innew{\I_n(m)\ =\ \lim\,\big(C^{}_1/C_1'\big)^n}
where the right-hand side is understood in the double scaling limit.

In order to evaluate $\I_n$ explicitly, and thereby relate $C_1$ and
$C_1',$ it suffices to choose $a$ such that
each side of Eq.~\RGreq\ is nonzero. A convenient such choice,
consistent with the ADHM constraints $\abar a=(\abar a)^T,$ is
$a'_{\alpha\dalpha}={\rm diag}(a'_{11\,\alpha\dalpha}\,,\cdots,
a'_{nn\,\alpha\dalpha})$ and $w_{k\,\alpha\dalpha}=0,$ in the notation of
Eq.~\bcanonical. For this choice, the eigenvectors of $\bigL$ are
the $n\times n$ antisymmetric matrices
$t_{ij},$ $1\le i<j\le n,$ defined by their matrix
elements $\big(t_{ij}\big)_{kl} = \delta_{ik}\delta_{jl}-
\delta_{il}\delta_{jk}\,$. From the definition of $\bigL$ (see Eq.~(3.10)
of (I)$\,$) one sees that $\bigL\cdot t_{ij}= |a'_{ii}-a'_{jj}|^2\cdot
t_{ij}\,$, so that the right-hand side of Eq.~\RGreq\ is simply
$(C_1)^n\cdot\prod_{\ijna}|a'_{ii}-a'_{jj}|^2\,$. On the other hand, the
left-hand side of Eq.~\RGreq\ is trivially evaluated for this choice of $a$.
The $\nu_i$ and $\N'_{ii}$ integrations are saturated solely by the mass term,
giving $(-2m\pi^2)^n$ and $(-m\pi^2)^n$, respectively, while the $\N'_{ij}$
integrations with $1\le i<j\le n$ are saturated by the \deltafcns,
and yield 
$\prod_{\ijna}|a'_{ii}-a'_{jj}|^2\,$. So the $a$ dependence indeed cancels
out as claimed, leaving
\eqn\Insolve{\I_n(m)\ =\ (2m^2\pi^4)^n\ .}
Equivalently, in the double scaling limit, from Eqs.~\Innew,
\Coneequals\ and \Conepequals:
\eqna\RGCC
$$\eqalignno{C_1\ &=\ \lim\,2m^2\pi^4C_1' \ ,&\RGCC a
\cr
\Lambda_{\sst N=1}^6\ &=\ \lim\,m^2 \Lambda_{\sst N=2}^4\ . &\RGCC b }$$

By identical arguments, one can also flow from the $N=4$ measure
\dmudefour\ to the $N=2$ measure \dmudeftwo, by inserting the $N=2$
invariant Higgsino mass term \dkmsix
\eqn\newflow{\exp\big(-2im\pi^2\Tr_n\,\M^{4T}(\Pinfty+1)\M^3\big)}
into the integrand of the former, and carrying out the integrations
over $\M^3$ and $\M^4$. This integration is proportional to $(\det\bigL)^2$
as required, and yields the relation
\eqna\RGCCnew
$$\eqalignno{C'_1\ &=\ \lim\,4m^4\pi^8C_1'' &\RGCCnew a}$$
or equivalently
$$\eqalignno{\Lambda_{\sst N=2}^4\ &=\ \lim\,m^4 \,
\exp(-8\pi^2/g^2_{\sst N=4})
 &\RGCCnew b }$$
in the double scaling limit. This result holds regardless of whether
the  expression \ifinclude\ is included in the
definition of the $N=4$ measure, as this term is subleading compared
to the mass term \newflow\ in this limit. 

Note that Eqs.~\RGCC{b} and \RGCCnew{b} are consistent with the standard
prescriptions in the literature for the RG matching of a low-energy and
a high-energy theory \refs{\RGorig,\RGinst}. The absence of numerical
factors on the right-hand sides of these relations reflects the
absence of threshold corrections in the PV scheme. 

\newsec{The classical $N=0$ measure}
Finally, we can flow from the $N=1$ measure to the purely bosonic
``$N=0$'' measure as well, by inserting a gaugino mass factor
\eqn\gauginomass{\exp
\big(-m\pi^2\,\Tr_n\,\M^{\gamma T}(\Pinfty+1)\M_\gamma\big)}
into the integrand of Eq.~\dmudef, and carrying out the $\M$ integration
using the above identities. In this way one finds for the $N=0$ measure
the $O(n)$ invariant expression:
\def\Czero{C^{\sst(0)}}
\eqn\dmudefzero{\eqalign{\int\dmuncl \ =\
&{(\Czero_1)^n\over\VolOn}\int\prod_{i=1}^nd^4w_i
\prod_{\ijn}d^4a'_{ij}\cr&\times\
\,\prod_{\ijna}\prod_{c=1,2,3}
\delta\big(\quarter\trtwo\tau^c[(\abar a)_{i,j}-
(\abar a)_{j,i}]\big)\, \det\bigL\ ,}}
where $\Czero_1$ is related to $C_1$ in the double scaling limit via
\eqn\RGCCnewest{\Czero_1\ =\ \lim\,2m^2\pi^4C_1\ =\
\lim\,2^{10}\,m^2\pi^4\,\Lambda_{\sst N=1}^6\ .}
 (In the 1-instanton case, the second line in Eq.~\dmudefzero\
is absent as always.)  
This expression may be seen to obey cluster decomposition by the same
arguments as for the $N=1$ and $N=2$ measures (see Secs.~2.4 and 3.4 of 
(I)$\,$);
in particular it is convenient for this purpose to represent $\det\bigL$
as a Grassmann integral analogous to the bosonic representation \detLrep\
for $(\det\bigL)^{-1}.$ As a check of Eq.~\dmudefzero, 
in the 2-instanton sector it can be shown to be equivalent to
 the first-principles (but $O(2)$-breaking) form for the
classical measure written down by Osborn \Osborn. The proof of this equivalence
is identical to that for the $N=1$  and $N=2$ cases considered in 
Secs.~2.5 and 3.5 of (I) and to
the $N=4$ case discussed above, and is left to the reader.

As stated earlier, Eq.~\dmudefzero\ is purely a classical
measure since it neglects the one-loop small-fluctuations 't Hooft determinants
over positive-frequency modes in the self-dual background. While such classical
collective coordinate integration measures have been studied before
in the ADHM problem, at the 2-instanton level \refs{\GMO,\Osborn},
they are more familiar in the context 
of BPS multi-monopoles \refs{\AH,\gm}.
There they correspond to  volume forms obtained by taking the
appropriate power of the classical 
metric \hbox{2-forms}. The classical hyper-K$\ddot{\rm a}$hler 
metric on the multi-monopole moduli 
space is physically important, as it governs the nonrelativistic scattering
of BPS monopoles in (3+1)-dimensions. 
In the ADHM case the corresponding classical
hyper-K$\ddot{\rm a}$hler metric is presently unknown (although the
classical volume form \dmudefzero\  may provide a useful
constraint). By analogy with
 monopoles, such a metric would govern the nonrelativistic scattering
of four-dimensional multi-instantons embedded in a \it five\rm-dimensional
space-time, a rather arcane physical problem. 

Assuming, instead, that one is primarily interested in the contribution of
ADHM multi-instantons to four-dimensional physics, knowledge of
the classical integration measure does not suffice. One must know
the one-loop  determinants as well, for two compelling reasons. First,
 these determinants  enter the semiclassical expansion  at order
$g^0,$ just like Eq.~\dmudefzero\ itself. Second, without them,
Green's functions turn out to be scale- and scheme-dependent, hence unphysical.
In particular, the classical 1-instanton factor $\Czero_1$ in the PV scheme
may be extracted from Eq.~(12.1) of \tHooft:
\def\muzero{\mu_0}
\eqn\Czeroequals{\Czero_1(\muzero)\ =\ 2^{10}\,\pi^4\,\muzero^8\,
\exp\big(-8\pi^2/g^2_{\sst N=0}(\muzero)\big)\ =\ 
 2^{10}\,\pi^4\,\muzero^8\,\big(\Lambda_{\sst N=0}/\muzero\big)^{22/3}}
which explicitly depends on the subtraction scale $\muzero$. (Presumably
Eq.~\RGCCnewest\ above
should be understood at the matching scale, $\muzero=m$.)
Only when the one-loop determinant is included, giving a contribution \tHooft
\eqn\oneloopdet{(\muzero\,|w|)^{-2/3}\,e^{-\alpha(1)}\ ,\quad
\alpha(1)\cong0.443307}
with $|w|$ the instanton size,
do the explicit factors of $\muzero$ in Eq.~\Czeroequals\
cancel out, leaving a scale-independent,
RG-invariant answer for the physical 1-instanton measure.
To date, 
 there has been substantial progress towards the
 calculation of these one-loop determinants 
in the general ADHM background
\refs{\detrefsO,\detrefsL}, 
based on the expressions for the
Green's functions  derived in \CGTone. 
Taken together with the classical measure
\dmudefzero\ above, complete
knowledge of these determinants would ultimately put ADHM multi-instantons
on the same solid field-theoretic
footing as  single instantons have been since the work
of 't Hooft. 

ND and TJH are supported by PPARC Fellowships; VVK by the TMR network
FMRX-CT96-0012; and MPM by the Department of Energy.
We thank Nick Manton and Fred Goldhaber for useful comments. 

\listrefs

\bye